\documentclass{aastex6}

\usepackage{amsmath}
\usepackage[version=3]{mhchem}

\newcommand{\wmm}{\,W\,m$^{-2}$}

\fullcollaborationName{}

\begin{document}

\title{Tutorial models of the climate and habitability of Proxima Centauri b: \\a thin atmosphere is sufficient to distribute heat given low stellar flux}

\author{Colin Goldblatt}
\affil{School of Earth and Ocean Sciences,\\
University of Victoria,\\
PO Box 1700 STN CSC, Victoria, British Columbia, V8W 2Y2, Canada.}

\begin{abstract}
Proxima Centauri b, an Earth-size planet in the habitable zone of our nearest stellar neighbour, has just been discovered. A theoretical framework of synchronously rotating planets, in which the risk of a runaway greenhouse on the sunlight side and atmospheric collapse on the reverse side are mutually ameliorated via heat transport is discussed. This is developed via simple (tutorial) models of the climate. These show that lower incident stellar flux means that less heat transport, so less atmospheric mass, is required. The incident stellar flux at Proxima Centauri b is indeed low, which may help enhance habitability if it has suffered some atmospheric loss or began with a low volatile inventory. 
\end{abstract}

\keywords{astrobiology --- radiative transfer ---planets and satellites: atmospheres --- stars: low-mass --- Earth}

\section{Introduction} \label{sec:intro}

Proxima b is a recently announced exoplanet \citep{Anglada-Escude2016}, presented as the first Earth-size planet to be found in the habitable zone. Proxima Centauri is the nearest star to Earth, 4.2 light years away. It is of M5.5V, with an effective temperature of 3050K and radius $0.14R_\text{Sun}$. Proxima b has $m\sin i = 1.3m_\text{Earth}$. It is on a 11.2 day orbit, so receives 65\% as much energy from its star as Earth does, and is thus in the ``habitable zone'' \citep{Anglada-Escude2016}. The importance of this discovery needs no further introduction. 

Many aspects of the habitability of Proxima b have been addressed by \citet{Ribas2016} and \citet{Turbet2016} (posted to the ArXiv contemporaneously with publication of the discovery) the latter focussing on climate. \citet{Turbet2016} use a General Circulation Model (GCM) to model potential climates. In this contribution, I take the contrasting but complementary approach of using simple, tutorial, climate models to examine the climate of Proxima b. This allows me to focus on intuition building for fundamental aspects of the climate system and to elucidate some (perhaps) important qualitative aspects relating to its habitability.  

I begin this paper with a discussion of the theory applicable to synchronously rotating planets around M-stars. Some of this is review, but I will also introduce some new theoretical framework, where I consider the risk of a runaway greenhouse on the sunlight side and the consequences of low stellar flux. 

Thereafter, I will use a simple model to show that, given the low stellar flux, much less redistribution of energy by atmospheric motion is required to maintain habitable conditions. Given that any atmosphere may be at risk of loss due to high XUV flux, this may substantially increase the chances that Proxima b is indeed habitable. 



\section{On the habitability of synchronously rotating planets around M-stars}

To have surface liquid water, a planet should receive neither too much nor too little energy from the star---so forms the basis of the ``circumstellar habitable zone''. Translating this into quantitative estimates of energy fluxes or circumstellar distances, boundaries outside which the existence of surface liquid water may be excluded, is thus a canonical problem in planetary climatology. The inner edge may be described by either a runaway water vapor greenhouse, which would bake the planet, or by massive loss of H (from water) to space, which would dessicate it. Venus, for example, probably experienced a runaway greenhouse followed by water loss. At the outer edge the fundamental limit is condensation of \ce{CO_2}, assumed to be the dominant greenhouse gas other than water. In the atmosphere, \ce{CO2} clouds would increase albedo (a positive feedback on cooling). If condensate were stable at the surface, atmospheric collapse would ensue. Given a low stellar flux and cool temperatures, the onset of this would likely be hastened by ice-albedo feedback causing global glaciation \citep[e.g.][]{abe-93,kwr-93}.

The runaway greenhouse is described further, as some theoretical development will follow. The amount of water in the atmosphere depends on temperature, but water is a greenhouse gas, so there is a positive feedback. With a deep column of water, the atmosphere becomes optically thick across the thermal infrared region. Only the atmosphere, not the surface, may then radiate to space and the temperature of the emission level asymptotes to a constant set by the thermodynamic and radiative properties of water. Consequently, there is a maximum amount of radiation that a moist atmosphere can emit \citep[282\wmm, ][]{Goldblatt2013}. If more sunlight than this is absorbed by an isolated column, or by the planet on average, then runaway surface warming occurs \citep[e.g.][]{simpson-27,nakajima-ea-92,Goldblatt2012}. Earth's tropics absorb more sunlight than the radiation limit. A runaway greenhouse is avoided for three reasons: (1) Heat is exported to the extratropics (2) Columns of unsaturated air allow more radiation out to space \citep{pierrehumbert-95} (3) deep convention initiates at around the surface temperature which corresponds to the radiation limit, increasing cloudiness and albedo.

Synchronously rotating planets were first though of as non-habitable as it was inferred that  the atmosphere would collapse on the dark side and trap all the volatiles there. However, pioneering work with an energy balance model \citep{Haberle1996} and later a general circulation models (GCM) \citep{Joshi1997} showed that energy redistribution by atmospheric circulation could prevent atmospheric collapse. Atmospheric heat transport is, to first approximation, proportional to surface pressure (atmospheric mass).
This has subsequently been elaborated with further GCM studies, usually focusing on planets receiving the same stellar flux as the modern Earth \citep{Joshi2003,Edson2011,Yang2013,Turbet2016}. An atmosphere with a surface pressure of 100\,hPa is though to be needed. 

The flip-side of the climate coin is that the sunlit hemisphere would be susceptible to a runaway greenhouse, but this has received less theoretical attention. The \citet{Haberle1996} energy balance model had a fixed greenhouse effect, independent of temperature, so necessarily avoided a runaway greenhouse: this likely helped frame the problem in terms of dark-side collapse only. Nonetheless, heat transport that prevents collapse would play the dominant role in avoiding a runaway greenhouse. Further, some of the most capable GCMs  have shown near total cloud cover around the substellar point \citep{Yang2013}.

Atmospheric thickness also pertains to habitability via Rayleigh scattering, pressure broadening and dilution of water. More atmosphere, so more Rayleigh scattering, cools a planet by increasing albedo at the blue end of the solar spectrum (Rayleigh scattering cross section is proportional to the reciprocal forth power of wavelength). Conversely, more atmosphere pressure broadens the absorption lines of greenhouse gases and warms the planet. The latter effect dominates for pressures of up to a few bars \citep{Goldblatt2009a}. Lastly, more atmosphere will dilute water vapour, making hydrogen escape less likely. 

The overall stellar flux received by Proxima b, 65\% of the modern solar constant, is lower than Earth received even at the start of the Sun's main sequence. The canonical problem of deep palaeoclimate is the Faint Young Sun Paradox \citep{sm-72}: despite low solar constant, how did Earth avoid ice-albedo feedback and persistent global glaciation? (Geologic evidence is that early Earth was ice free more often than not \citep[e.g.][]{nisbet-87}, though there is no agreement on the climate forcings that allowed for that). But would Proxima b not be pan-glacial?

The spectrum of M-stars is shifted far to the red relative to the Sun, affecting climate and habitability. 
Snow and ice are near white around the Wein peak of sunlight, but dark at the Wein peak of M-star emission, so the ice-albedo feedback disappears \citep{Joshi2012} (also, on a synchronous rotator, snow would fall on the unilluminated side). 
Little blue light is received, so Rayleigh scattering is of little importance for pressures of up to a few bar. 
Both of these effects bode well for weakly illuminated Proxima b: cooling mechanisms which endanger a planet near the outer edge of the habitable zone do not apply. 

However, Proxima Centurai is a flare star, with a ratio of XUV flux to bolometric luminosity $\sim60$ times higher than the Sun \citep{Ribas2016}. Whereas the overall energy flux the planet receives determines climate the XUV flux will heat the thermosphere and drive atmospheric escape. The risk of substantive atmospheric loss is a real \citep[see extensive discussion in ][, and references therein]{Ribas2016}. 

Also pertaining to atmospheric mass, the initial volatile inventory of M-star habitable zone planets may be low \citep[][though again see discussion in \citet{Ribas2016}]{Lissauer2007}.

The climate and habitability problem for Proxima b may be summarised. The planet may well be synchronously rotating, so is exposed to twin perils of a runaway greenhouse of the sunlight side and atmospheric collapse on the dark side. These can be avoided by advective heat transport from front to back, which depends on atmospheric mass. However, Proxima b may have had a low initial volatile inventory or suffered atmospheric loss. Overall stellar flux is low. The question may be posed as: how little atmosphere would be sufficient to transfer enough energy---and how does this depend on incident stellar flux? 

\section{Climate with no greenhouse effect or heat transport}

The simplest case is to consider a planet with no atmosphere (or an atmosphere with no absorbers, and no heat transport), which is an elementary problem in climate modelling. This provides insight into the climate forcing of M-star planets under different insolation, and additionally provides a reference case to compare subsequent models with an atmosphere. 

\subsection{Model description}

The surface of the planet is in energy balance, with absorbed stellar radiation and a geothermal heat flux balanced by thermal emission, which is taken to be a black body. Thus:
\begin{equation}
F_s + F_g = \sigma T_\text{eff}^4,
\end{equation}
with 
\begin{equation}
F_s = 
\begin{cases}
    S(1-\alpha)\cos \phi  & \text{for } \phi < \tfrac{\pi}{2} \\
    0  & \text{for } \phi \geq \tfrac{\pi}{2}
\end{cases}
\end{equation}
where $F_s$ is the stellar flux incident per unit surface area, solar (or stellar) constant $S$ is the flux at the top of the atmosphere at the substellar point (modern Earth, $S_0 = 1368$\wmm), $\alpha$ is the bond albedo, $\phi$ is the angular separation from the substellar point, $\sigma = 5.67\times 10^{-8}$\wmm\,K$^{-4}$ is Stefan's constant and $T_\text{eff}$ is the effective temperature. Geothermal heat flux $F_g$ may be set via comparative planetology; $F_{g,\text{Earth}} = 0.08$\wmm, dominated by radiogenic heat production in the mantle and internal cooling, $F_{g,\text{Io}} = 2$\wmm, dominated by tidal heating. A conservative value of $F_g = 0.2$\wmm{} is used. 

Note that surface temperature for the airless case is equivalent to effective temperature $T_\text{eff}$. The meteorological definition is required; effective temperature is the emission temperature required for a column to be in energy balance with $F_s$, noting that \emph{planetary albedo must be considered}.

\subsection{Results}

\begin{figure}
\begin{center}
\includegraphics[width=0.7\textwidth]{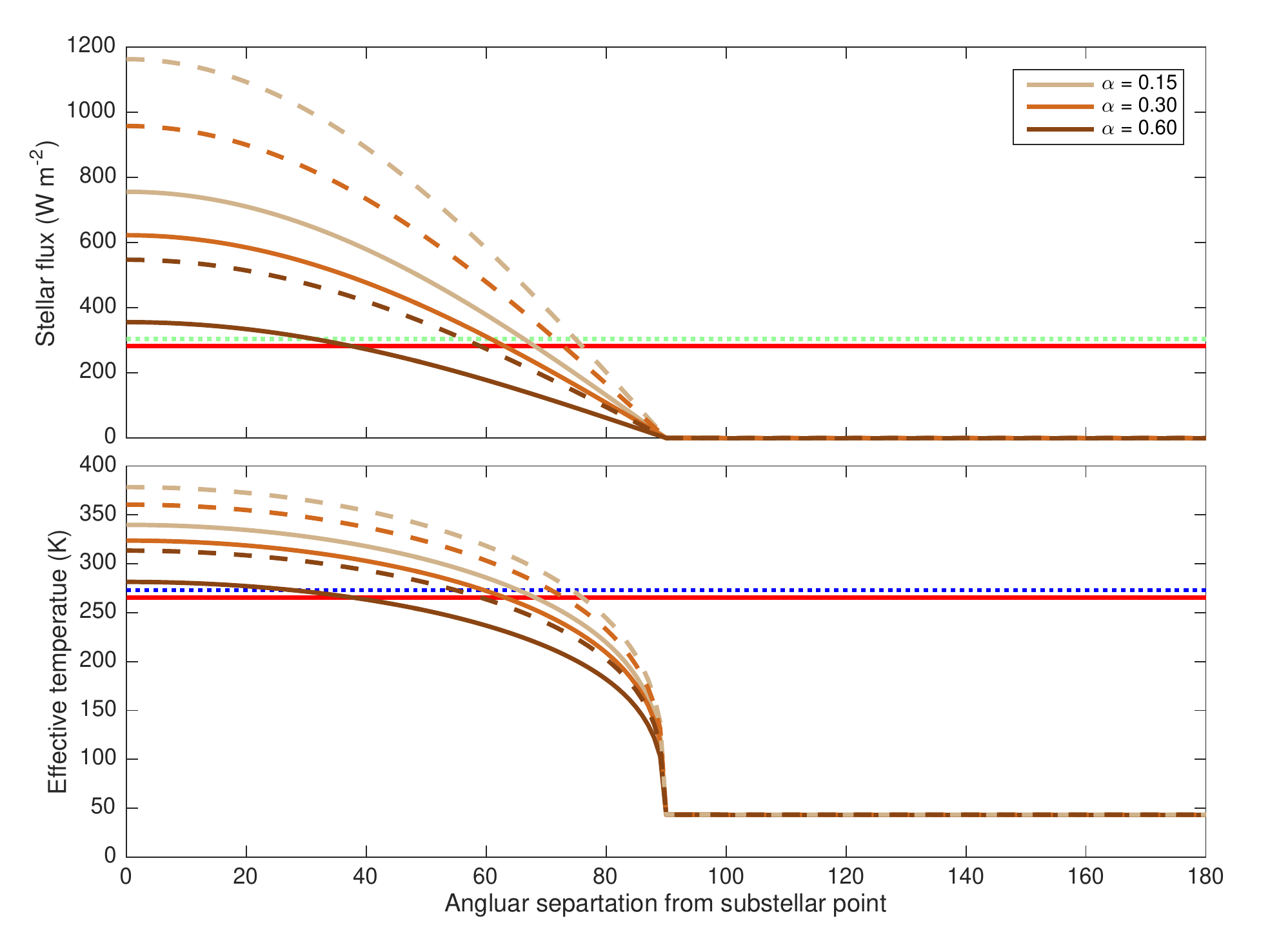} 
\end{center}
\caption{Incident stellar flux (top) and effective temperature / surface temperature for airless case (bottom). Solid lines for $S = 0.65S_0$, dashed line for$S = S_0$. Horizontal lines are reference fluxes and temperatures: solid red is the radiation limit for runaway greenhouses, dotted green is the diurnal average insolation at Earth's equator, blue is the triple point temperature for water. }
\end{figure}

Incident sunlight and surface temperature for the airless planet are shown in Figure 1 for the stellar flux at Proxima b, $S = 0.65S_0$ and  the reference value $S = S_0$. Representative albedos used are 0.15 (bare rock), 0.3 (patch water clouds, as modern Earth) and 0.6 (complete water cloud coverage). 

Incident stellar flux exceeds the runaway greenhouse threshold over much of the sunlit hemisphere. 
This illustrates one bound on habitable climate for synchronously rotating planets; with surface water, and were the light and dark hemispheres thermally isolated, the illuminated hemisphere would heat beyond the stability of liquid water.
The difference between $S = 0.65S_0$ and $S = S_0$ is immediately apparent. With modern Earth insolation, most of the sunlit hemisphere would be unable to radiate the absorbed sunlight locally, so prodigious heat export to the dark hemisphere would be required. With the stellar flux at Proxima b, the excess sunlight is much smaller. The implication is clear: given that stellar flux at Proxima b is low, less energy redistribution is required and, by inference, less atmosphere is needed. 

Surface temperature for the airless case is instructive too: there is a band which is both sunlight and temperate. One could imagine locally stable soil moisture and life within the soil. 

\section{Climate with an atmosphere}

\subsection{Model description}

I modify the energy balance model of \citet{Haberle1996} by trivial addition of a geothermal heat flux and substantive addition of a temperature dependent greenhouse effect, corresponding to the water vapour feedback on climate and ultimately a runaway greenhouse. The original model solves for energy balance at the surface and in a single-layer atmosphere on the both the light and dark sides of the planet. The atmosphere is transparent to sunlight and grey to thermal radiation. The two sides are connected by an atmospheric heat flux \citep[see Figure 1 of ][]{Haberle1996}. The equations (not written out in the original paper) are:
\begin{eqnarray}
\mathrm{Light, surface:} \quad 0 & = & \tfrac{S}{2}(1-\alpha) + \varepsilon\sigma T_{la}^4 - \sigma T_{ls}^4 \\
\mathrm{Light, atmos.:} \quad 0 & = & -2\varepsilon\sigma T_{la}^4 +\varepsilon\sigma T_{ls}^4  - A(T_{la}-T_{da})\\
\mathrm{Dark, surface:} \quad 0 & = &   \varepsilon\sigma T_{da}^4 - \varepsilon\sigma T_{ds}^4 \\
\mathrm{Dark, atmos.:} \quad 0 & = & -2\varepsilon\sigma T_{da}^4 + \varepsilon\sigma T_{ds}^4 + A(T_{la}-T_{da})
\end{eqnarray}
$T$ is temperature, with subscripts $l$ for light, $d$ for dark, $s$ for surface and $a$ for atmosphere. Emissivity $0<\varepsilon<1$ is constant and equals absorptivity. The advection parameter is 
\begin{equation}
A = \frac{p_s c_p}{g t_\text{adv}}
\end{equation}
with advective timescale
\begin{equation}
t_\text{adv} = \frac{L}{U}. 
\end{equation}
Using modern Earth values for all variable in $A$ (surface pressure $p_s = 10^5$\,Pa, specific heat capacity of air $c_p = 1000$\,J\,kg$^{-1}$\,s$^{-1}$, gravity $g = 9.8$\,m\,s$^{-2}$, length scale $L = \frac{\pi}{2}r$, planet radius $r = 6.4\times 10^{6}$\,m and windspeed $U = 10$\,m\,s$^{-1}$) gives $A_0 = 10$\wmm\,K$^{-1}$. 

My modified equations are:
\begin{eqnarray}
\mathrm{Light, surface:} \quad 0 & = & \tfrac{S}{2}(1-\alpha) + \varepsilon_l\sigma T_{la}^4 - \sigma T_{ls}^4 + F_g\\
\mathrm{Light, atmos.:} \quad 0 & = & -\varepsilon_l\sigma(T_{la}^4 + T_{lr}^4) + \varepsilon_l\sigma T_{ls}^4  - A(T_{la}-T_{da})\\
\mathrm{Dark, surface:} \quad 0 & = &   \varepsilon_d\sigma T_{da}^4 - \sigma T_{ds}^4 + F_g \\
\mathrm{Dark, atmos.:} \quad 0 & = & -\varepsilon_d\sigma(T_{da}^4 + T_{dr}^4) + \varepsilon_d\sigma T_{ds}^4 + A(T_{la}-T_{da})
\end{eqnarray}
where $F_g$ is a geothermal heat flux and $T_r$ is the temperature of atmospheric radiation to space (defined below). Emissivity is no longer a free parameter, but set as 
\begin{equation}
\varepsilon = 1 - e^{-\tau}
\end{equation} 
where optical depth comprises contributions from water vapour and other greenhouse gases
\begin{equation}
\tau = \tau_\text{H2O} + \tau_\text{dry}.
\end{equation}
I take $\tau_\text{dry} = 0.5$, somewhat similar to Earth's atmosphere. Optical depth from water is directly proportional to a representative saturation vapor pressure;
\begin{equation}
\tau_\text{H2O} = kp_\text{sat}
\end{equation}
with absorption coefficient $k = 0.001$. For convenience, saturation vapor pressure is approximated \citep{nakajima-ea-92} as
\begin{equation}
p_\text{sat} = p_{\text{sat},0}\exp\left( \frac{L_0}{RT}\right)
\end{equation}
with $p_{\text{sat},0} = 1.4\times 10^{11}$\,Pa, $L_0 = 43655$\,J\,mol$^{-1}$ and $R = 8.14$\,J\,mol$^{-1}$\,K$^{-1}$. For the light side, $T_{ls}$ is appropriate for calculation of $p_\mathrm{sat}$, for this will represent the column water vapour. For the night side, however, there is commonly a temperature inversion (the atmosphere is warmer than the surface), so the maximum of $p_\mathrm{sat}(T_{ds})$ or $0.9p_\mathrm{sat}(T_{da})$ is used (the factor of 0.9 is based on an assumption that the inversion level is at nine-tenths of the surface pressure). 

Specification of atmospheric radiative temperature, $T_r$, is key to representing the runaway greenhouse. When the atmosphere is not optically thick from water, the radiative temperature will be the bulk atmospheric temperature $T_a$. When the atmosphere is optically thick with water, radiation is emitted only from the level where  $\tau_\text{H2O} \sim 1$ \citep{simpson-27,nakajima-ea-92,Goldblatt2012}. Thus I set 
\begin{equation}
T_r = 
\begin{cases}
    (1-\varepsilon_\text{H2O})T_a + \varepsilon_\text{H2O}T_\text{limit},& \text{for } T_a < T_x\\
     (1-\varepsilon_\text{H2O})T_a + \varepsilon_\text{H2O}T_\text{limit} + 0.01(T_a-T_x)^2,              & \text{for } T_a \geq T_x
\end{cases}
\end{equation}
where $T_\text{limit} = 265$\,K is the blackbody temperature corresponding to a limiting flux of 282\wmm{} \citep{Goldblatt2013}. This parameterization, with parameters tuned as above, well represents modern radiative transfer model output \citep{Goldblatt2013}.

At very high surface temperatures ($T>1500$\,K) the atmosphere becomes sufficiently hot aloft that radiation escapes through the 4\micron{} water vapor window, allowing a hot dry equilibrium temperature to be found \citep{Goldblatt2015}. This is difficult to represent well with a single layer atmosphere; it is crudely parametrized by setting some maximum atmospheric temperature $T_x = 600$\,K, above which the radiative temperature again increases. 

With $k = 0$, my modified model simplifies to the original \citep{Haberle1996} model. 

\subsection{Results}

\begin{figure}
\begin{center}
\includegraphics[width=0.7\textwidth]{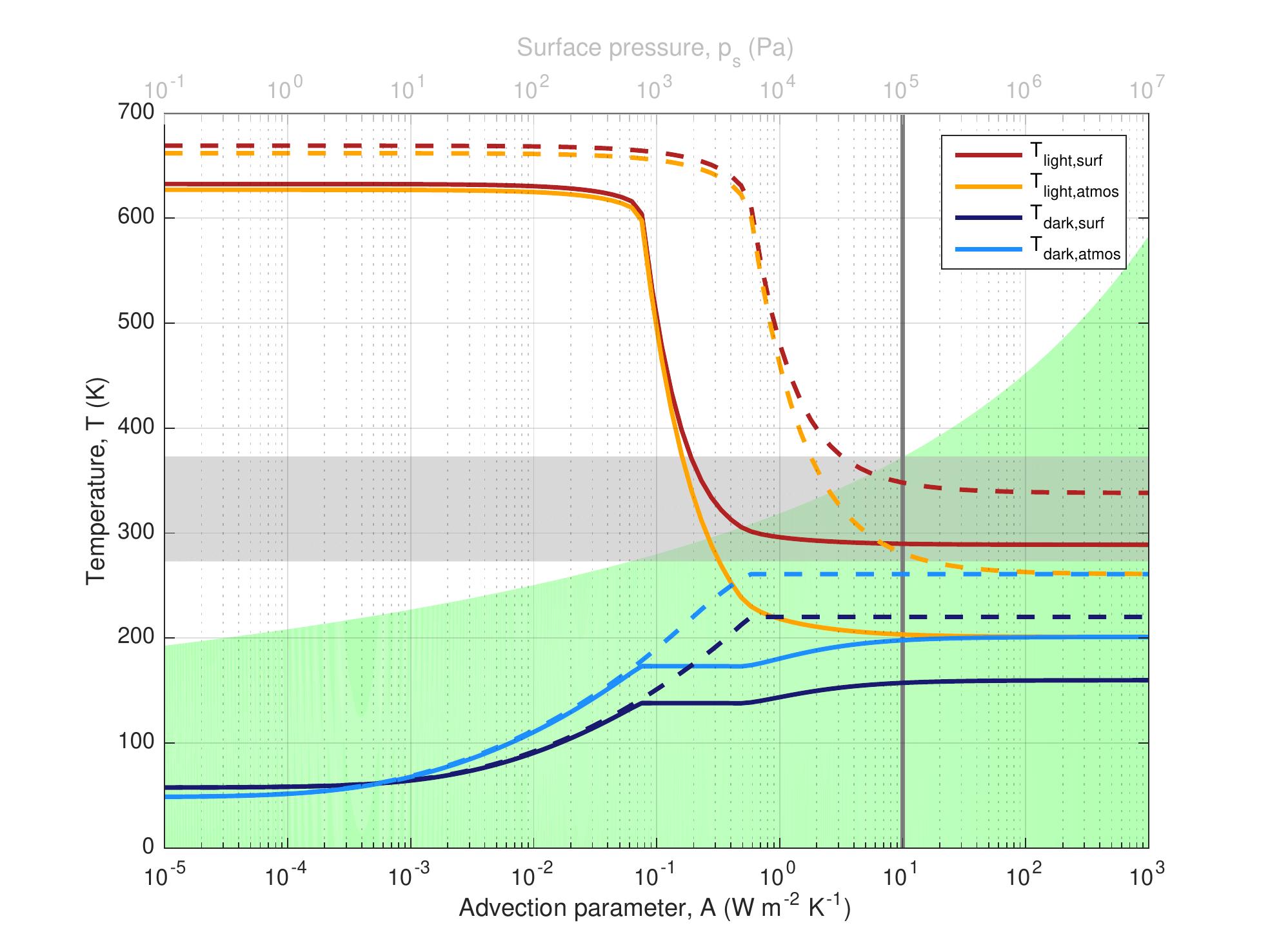} 
\end{center}
\caption{Model temperatures (solid lines for $S = 0.65S_0$, dashed line for$S = S_0$). Light grey shaded areas are modest surface temperatures. Green shaded area corresponds to water vapour mixing ratios less than 0.5. Vertical grey line is the reference advection parameter, $A_0 = $\wmm\,K$^{-1}$. Note that pressure scales linearly with $A$. }
\end{figure}

Surface and atmosphere temperatures for light and dark side of a synchronously rotating planet, for $S = S_0$ and $S = 0.65S_0$, are shown in Figure 2. These serve as the basis for assessing how much atmosphere is required for habitability via three criteria: (1) moderate surface temperatures on the light side, for mesophile life. (2) water vapour not the dominant atmospheric constituent, to avoid rapid water loss. (3) Sufficiently warm temperatures on the night side to avoid atmospheric collapse. 

Moderate temperatures are sketched with a representative temperature range of $273<T_s<373$\,K (grey area). Avoidance of a moist atmosphere is sketched with as $p_s > p_\text{sat}(T)$, equivalently water vapour mixing ratio $x_\text{H2O} < 0.5$ (green area). The intersection of green and grey areas for day side surface temperature is the locus of habitable conditions. With $S = 0.65S_0$, the planet is in this region with $A$ an order of magnitude lower than if $S = S_0$. Recall that $A \propto p_s$; the implication is that a reduced solar constant will allow a habitable day side with an order of magnitude less atmosphere.

Atmospheric collapse can be assessed by comparing night side surface temperatures (which are colder than the atmosphere) to the thermodynamic properties of popular atmospheric gases.
For $S = 0.65S_0$, the low $A$ end of day side habitability corresponds to $T_{ds} = 138$\,K.
Hydrogen and nitrogen have critical point temperatures of 33\,K and 126\,K, so these would have no condensed phase and are absolutely safe as bulk atmospheric gases. The triple and critical points of methane are 91\,K and 191\,K. At 138\,K, $p_{sat,CH4} = 5.8\times10^5$\,Pa, so a thick methane atmosphere is reasonable. Comparative planetology supports this; methane is an important (though condensible) atmospheric constituent on Titan where $p_s = 90$\,K. The triple point for carbon dioxide is 217\,K, so the situation here is more delicate. The saturation vapour pressure is plotted as a function of night side surface temperature (Figure 3). At 138\,K,  $p_{sat,CO2} = 135$\,Pa, several times higher than modern Earth p\ce{CO2} of 40\,Pa. The corresponding \ce{CO2} mixing ratio would be 3 -- 20\%: carbon dioxide would be stable as a minor atmospheric constituent. The triple point of water is 273\,K; ice formation on the dark side is inevitable. With a very low water inventory, the risk of cold trapping all the water exists. With a medium-size inventory, glaciers would flow to the terminator, melt and evaporate. With a lot of water, there would be a global ocean like Earth's, and water distribution would not be a problem. 

\begin{figure}
\begin{center}
\includegraphics[width=0.7\textwidth]{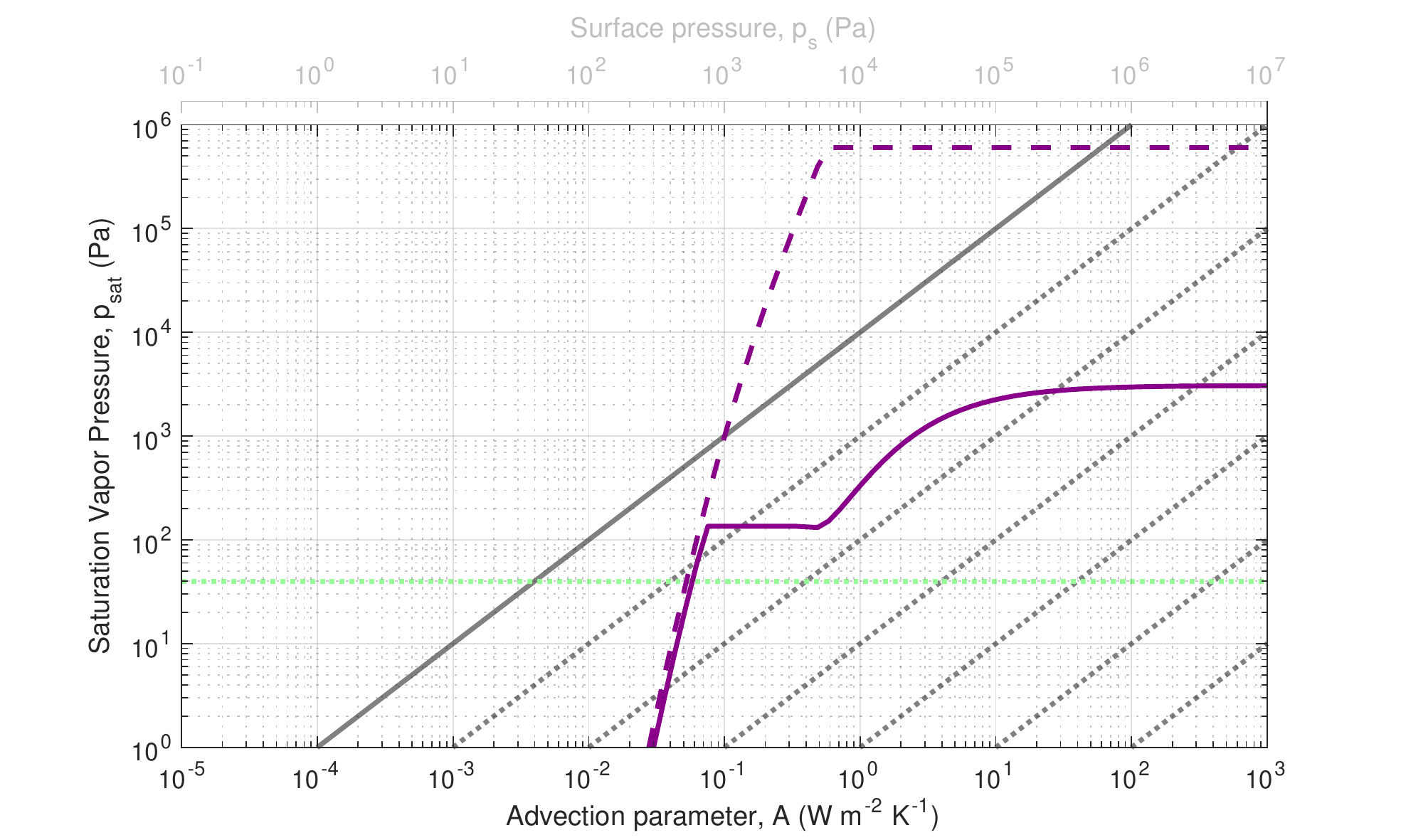} 
\end{center}
\caption{Saturation vapor pressure of \ce{CO2} corresponding to $T_{ds}$ (solid line $S = 0.65S_0$, dashed line $S = S_0$). p\ce{CO2}$> p_\text{sat,CO2}$ would cause deposition and low partial pressure implies atmospheric collapse. Grey solid line indicates a pure \ce{CO2} atmosphere (using $p_s$ implied by $A$) and dotted grey lines mixing ratios of 0.1 to $10^{-6}$. The green dotted line is the partial pressure of \ce{CO2} on Earth in 2016. Note that pressure scales linearly with $A$.}
\end{figure}

\section{Discussion} \label{sec:disc}

Simple climate models are some of the best teachers. Here, the the theoretical framework is risk of runaway greenhouse on the sunlight hemisphere and collapse on the reverse. Models clearly show the extent of the potential runaway greenhouse problem depending on incident stellar flux, and the amount of atmospheric heat transport required to offset this decreasing as incident stellar flux decreases. For Proxima b, low stellar flux means that the requirements on the atmosphere are less stringent. To achieve the goal of a temperate climate, an atmosphere of mostly nitrogen with minor \ce{CO2} seems the best choice.

So far, the atmospheric composition has been set as a contrivance of the modeller. This is, of course, not so: real terrestrial planet atmospheres are controlled by a mix of geological and biological processes. Indeed, the proposal of \emph{life detection by atmospheric analysis} \citep{lov-65} arose from the observation of biological control on Earth's atmosphere, indeed the proposer of this saw Earth's atmosphere as a biological contrivance \citep{lov-72,Lovelock1974}. Contingency in atmospheric evolution means that there is a paradox: habitability and inhabitance are inseparable \citep{Goldblatt2016}. 

Theory on the control of Proxima b's atmospheric composition must await a later contribution; differences relative to solar system planets are expected. 
For example, \ce{CO2} will likely condense to some extent (Mars-type control), but in if there is a water ocean then carbonate deposition would be expected (Earth-style control)---what would the dynamics of these contributions be? 
The most wonderful thing about Proxima b is, of course, that we will likely be able to characterize his atmosphere---its presence or absence,  its temperature and composition---in my lifetime, and thereby prove all our theories wrong. 

\acknowledgments
Financial support came from an NSERC discovery grant.


\end{document}